# Tests of sunspot number sequences:
# 4. Discontinuities around 1946 in various sunspot number and sunspot group number reconstructions

M. Lockwood[1] • M.J. Owens[1] • L. Barnard[1]

**Abstract**. We use five test data series to search for, and quantify, putative discontinuities around 1946 in five different annual-mean sunspot number or sunspot group number data sequences.  The data series tested are: the original and new versions of the Wolf/Zurich/International sunspot number composite [$R_{ISNv1}$ and $R_{ISNv2}$, respectively] (Clette *et al.*, 2007, 2015); the corrected version of $R_{ISNv1}$ proposed by Lockwood *et al.* (2014a) [$R_C$]; the new "backbone" group number composite proposed by Svalgaard and Schatten (2016) [$R_{BB}$]; and the new group number composite derived by Usoskin *et al.* (2016) [$R_{UEA}$]. The test data series used are: the group number $N_G$ and total sunspot area $A_G$ from the Royal Observatory, Greenwich / Royal Greenwich Observatory (RGO) photoheliographic data; the CaK index from the Mount Wilson Observatory (MWO) spectroheliograms in the Calcium II K ion line; the sunspot group number from the MWO sunspot drawings, $N_{MWO}$; and the dayside ionospheric F2-region critical frequencies measured by the Slough ionosonde, foF2. These test data all vary in close association with sunspot numbers, in some cases non-linearly.  The tests are carried out using both the "before-and-after" fit-residual comparison method and the correlation method of Lockwood *et al.* (2014a), applied to annual mean data for intervals iterated to minimise errors and to eliminate uncertainties associated with the precise date of the putative discontinuity. It is not assumed that the correction required is by a constant factor, nor even linear in sunspot number.  It is shown that a non-linear correction is required by $R_C$, $R_{BB}$ and $R_{ISNv1}$, but not by $R_{ISNv2}$ or $R_{UEA}$. The five test datasets give very similar results in all cases. By multiplying the probability distribution functions together we obtain the optimum correction for each sunspot dataset that must be applied to pre-discontinuity data to make them consistent with the post-discontinuity data. It is shown that, on average, values for 1932-1943 are too small (relative to later values) by about 12.3% for $R_{ISNv1}$ but are too large for $R_{ISNv2}$ and $R_{BB}$ by 3.8% and 5.2%, respectively.  The correction that was applied to generate $R_C$ from $R_{ISNv1}$ reduces this average factor to 0.5% but does not remove the non-linear variation with the test data, and other errors remain uncorrected.  A valuable test of the procedures used is provided by $R_{UEA}$, which is identical to the RGO $N_G$ values over the interval employed.

# 1. Introduction

The sunspot group number [$R_G$] was introduced by Hoyt and Schatten (1994, 1998). For after about 1900 it matches quite well the behaviour of sunspot numbers, such as version 1 of the Wolf/Zurich/International sunspot number composite [$R_{ISNv1}$] (Clette et al., 2007) but is well known to be significantly lower for earlier years (e.g. Lockwood et al., 2014b). This journal special issue includes two articles detailing two new sunspot group number series that are intended to be homogeneous and of stable calibration. Svalgaard and Schatten (2016) have proposed the "backbone" sunspot group series [$R_{BB}$] and Usoskin et al. (2015) propose a group number series that is here termed $R_{UEA}$. Compared to the (suitably-scaled) original $R_G$, both these new group number data series give higher values before 1900, but in the case of $R_{BB}$ they are radically higher. The major differences between $R_{BB}$ and $R_{UEA}$ arise from the method used to calibrate the historic data. The backbone series passes the calibration from one dataset to an adjacent one using a relationship between the two (usually a regression fit for the period of overlap between the two). This is called "daisy-chaining" and the problem with this method is that both systematic and random errors, compared to modern values, compound as one goes back in time. Furthermore, as discussed in Article 3 of this series (Lockwood et al., 2015c), there are problems and pitfalls with regression techniques in general, and there are concerns about the way that they were implemented by Svalgaard and Schatten (2016) in the generation of $R_{BB}$ (specifically, the assumption that data from different observers are proportional to each other is not generally correct in either principle or practice). Usoskin et al. (2016) avoid all of these pitfalls, and the potential for error propagation inherent in daisy-chaining, by devising a method that calibrates all data against one standard dataset. Note that, in general, observed group numbers from different observers vary non-linearly (Usoskin et al., 2016; Lockwood et al., 2016c).

In addition to these new group number series, a new version of the Wolf/Zurich/International sunspot number composite (ISN version 2, $R_{ISNv2}$) has recently been issued by Solar Influences Data Analysis Center (SIDC, the Solar Physics research department of the Royal Observatory of Belgium). Like $R_{BB}$, this uses daisy-chaining of calibrations and, also like $R_{BB}$, gives larger values for the 18[th] and 19[th] centuries (Clette et al., 2015). A less "root-and-branch" approach to correcting $R_{ISNv1}$ was taken by Lockwood et al. (2014a) who made simple corrections for known errors to generate a "corrected" series, $R_C$. It should be noted that because $R_C$ makes only corrections at two dates in the series, other errors in $R_{ISNv1}$, such as the error in modern data due to the drift in the Locarno standard (Clette et al., 2015), are carried forward and not corrected.

This paper concentrates on differences between these sunspot number and sunspot group number data series in the 20[th] century, specifically around 1946. Larger differences, inferred from geomagnetic activity data, low-latitude auroral sightings and cosmsogenic isotope abundances in ice sheets, tree trunks and meteorites, are found for earlier years which are discussed in Article 2 (Lockwood et al., 2016b) and in the paper by Asvestari et al. (2016). Changes around 1946 are of interest as there has been discussion about a putative inhomogeneity in the calibration of the original Zürich sunspot number data series [$R_{ISNv1}$] that has been termed the "Waldmeier discontinuity", as discussed in Article 1 (Lockwood et al., 2016a). This is thought to have been caused by the introduction of a weighting scheme for sunspot counts according to their size, and a change in the procedure used to define a group and, in particular, the "evolutionary" aspect of the new sunspot group classification scheme (called the Zürich scheme) introduced by Waldmeier (Waldmeier, 1947; Kiepenheuer, 1953). This raises two important questions. (1) What is this the correct quantification of this effect? (2) Which datasets employed the Zürich classification scheme and so would be subject to any such effect or may have been re-calibrated using the Zürich data? It is now agreed that $R_{ISNv1}$ needs correcting for this effect but it is unclear if, why and how it influences other data series. Tests comparing against ionospheric data (Lockwood et al., 2015a), auroral sightings and geomagnetic data (Lockwood et al., 2015b) all suggest that, somehow, an excessive allowance for the Waldmeier discontinuity has also been introduced into $R_{BB}$.

Often in the past, corrections to sunspot numbers have often been applied by taking ratios, which implicitly assumes that proportionality between the different data applies. This is often not the case (Lockwood et al, 2016c). A particular problem occurs when sunspot numbers are small because the errors in such ratios become highly asymmetric and both the ratio and its error tend to infinity if the denominator approaches zero. Two ways of avoiding this (in its most extreme form) have been employed. The first is to consider ratios only when the denominator exceeds an arbitrarily-chosen threshold, but this preferentially removes sunspot minimum values which do not always go to zero. The second way is to employ averages over one or more solar cycles so that the denominator remains large (outside grand minima): this matches long-term average values but loses all information about cycle amplitudes. Consequently, Lockwood et al. (2014a) devised two different procedures to test for discontinuities. The first fits the same polynomial form of a proxy or test dataset to two intervals, one before the putative error, one after it, and studies the probability of the difference in the mean fit residual for the "before" and "after" intervals. The second method looks at the effect of various assumed discontinuities on the correlation between the data and the test data. Generally the methods provide similar answers but uncertainties are lower for the fit-residual procedure, so it is the more stringent test. We here make a number of improvements to the implementation of the Lockwood et al. (2014a) methods.

In the original analysis of the Waldmeier discontinuity by Svalgaard (2011), it was assumed that the correction required was a single multiplicative ("inflation") scaling factor $f_R$, such that before the discontinuity the data were adjusted by multiplying by $f_R$ (i.e., the corrected sunspot number is $R' = f_R R$). This assumption was also used by Lockwood et al. (2014a) and Lockwood et al. (2016a). In general, it is not clear what the functional form of the correction for the Waldmeier discontinuity should be and it will be different for different sunspot number and sunspot group number series, depending on how they were compiled. Svalgaard et al. (2016) and Clette and Lefèvre (2016) have analysed the effect on Zürich sunspot numbers by applying both the pre-1946 and post-1946 procedures to modern data. The effects depend on timescale and, in general, are non-linear in $R$. The effect on annual averages is not as clear as for daily or monthly means.

We here generalise the correction by allowing for a both a zero-level offset $\delta$ and a nonlinear dependence (with an exponent $n$ of $R$) as well as the scaling factor [$f_R$] and a zero-level offset $\delta$. The exponent $n$ would be unity for a linear correction (i.e. the correction required is the same at all $R$): note that $n$ could be either greater or smaller than unity. For a proportional correction $n = 1$ and $\delta = 0$. We apply the correction to the data before the putative discontinuity. Hence, the corrected $R$ [$R'$] for a discontinuity at a date $t = t_d$ is defined by:

$$R' = R \qquad \text{for } t \geq t_d$$

$$R' = f_R R^n + \delta \qquad \text{for } t < t_d \qquad (1)$$

Lockwood et al. (2014a) used all of the sunspot group number dataset from the Royal Observatory Greenwich / Royal Greenwich Observatory (hereafter "RGO") which covers the years 1875 to 1976. The stability of the calibration of the earliest of these data (before 1885) has been questioned (Cliver and Ling, 2016) and this may have influenced the derived correction (Clette and Lefèvre, 2016). In the present paper, as in Lockwood et al. (2016a), we avoid using any RGO data from before 1900.

In addition, Clette and Lefèvre (2016) make the valuable point that there are other factors which may have influenced the correction factor derived by Lockwood et al. (2014a). The first is that other errors in the data series may be influencing the optimum correction for Waldmeier discontinuity. The second is that the precise date of the discontinuity [$t_d$] has an effect and is not known because Waldmeier's documentation is not clear on when the changes were actually implemented. Clette and Lefèvre (2016) make use of the ratio of $R/R_G$ to define $t_d$. This had been avoided by Lockwood et al. (2014a) because the error in such ratios tends to infinity when $R_G$ tends to zero and $R_G$ has a minimum in 1944, just before the putative discontinuity: hence changes would naturally become more apparent as sunspots began to rise in the subsequent cycle. From the $R/R_G$ ratio, Clette and Lefèvre (2016) place the discontinuity in 1946 (whereas Lockwood et al (2014a, 2016a) used 1945), although they note that there is some documentary evidence that at least some of the new procedures that are thought to be the cause of the discontinuity were in use earlier than this date. Clette and Lefèvre (2016) analysed the effects of both the start date of the comparison and the assumed discontinuity date [$t_d$] on the $R_{ISNv1}$ correction derived by Lockwood et al (2014a). They reproduced the Lockwood et al. (2014a) values when using the same dates; however, they found that the required correction could be larger if other dates were adopted. The analysis presented in this paper makes improvements to the procedure of Lockwood et al. (2014a) to remove these potential uncertainties.

## 2. Analysis

The analysis presented here employs five "test" data series and is applied to five "tested" sunspot reconstructions.

*2.i Tested Sunspot Data Series*

We here test how five different sunspot number or sunspot group number data series behave around 1946: these are summarised in Table 1 and compared in figure 1.

1. The original composite of Wolf/Zürich/International sunspot number [$R_{ISNv1}$] which is still available in the archive section of the SIDC website, but has not been updated since 1$^{st}$ July 2015. This is a composite of sunspot numbers, initially generated by Wolf and continued at the Zürich observatory until 1980 and then subsequently compiled by SIDC (until July 2015 when it was replaced by version 2). This is the dataset that moved to the Zürich classification scheme and so will show all aspects of the Waldmeier discontinuity. As for all the tested data series, with the exception of that by Usoskin et al (2016), the calibration is by daisy-chaining, i.e. the calibration is passed from one observer to the next (or previous) one by comparison of simultaneous data from both observers.

2. The new SIDC composite of Wolf/Zürich/International sunspot number, $R_{ISNv2}$, which became SIDC's default series on 1$^{st}$ July 2015. This corrects for a number of causes of long-term change in $R_{ISNv1}$, including the Waldmeier discontinuity and the correction of a drift in the calibration of the main station (Locarno) which had varied by ±15% between 1987 and 2009 (Clette et al., 2015). Note that this no longer uses the traditional scaling factor of 0.6 employed in $R_{ISNv1}$.

3. The new "Backbone" group sunspot number $R_{BB}$ proposed by Svalgaard and Schatten (2015). This differs in its long-term variation from the Hoyt and Schatten (1994, 1998) group number, $R_G$, and dispenses with the scaling factor of 12.08 introduced by *Hoyt and Schatten* (1994, 1998) (to make means of $R_G$ and $R_{ISNv1}$ the same in modern data). $R_{BB}$ is the mean of the results of two different methods: taking such a mean has the problem that, although errors can be halved, any error in either method is propagated into the final result, something that can be avoided if a probabilistic combination technique is applied. The first method involves daisy chaining of compiled "backbone" data sequences using linear regression (except for the earliest join when a different method is used: however this is not within the interval studied in the present paper and so this inhomogeneity in the series compilation is not a factor for this paper). The assembly of the backbones assumes proportionality and although their use reduces the number of linear regressions between backbones, it makes no difference to the number of observers through which the calibration is passed in the daisy chaining. The second method involves taking the largest group number defined by any observer in each year and scaling this to a backbone series. That four such intervals are required implies the relationship of the largest value to the optimum values changes over time and the calibration of this is again passed from one sequence to the previous one and hence this is also daisy-chaining. The daisy-chaining calibrations in $R_{BB}$ assume not only linearity of the data between different observers but also proportionality, which is not in principle correct and generated errors in the tests carried out by Lockwood et al. (2016c).

4. The "corrected" sunspot number" $R_C$ proposed by Lockwood et al. (2014) which is $R_{ISNv1}$ with their best estimate of the correction required for the Waldmeier discontinuity (plus, for earlier times, the correction of derived by *Leussu et al.* [2013] using data by Schwabe which applies to all data before 1848). This series contains no correction for any other errors, such as the Locarno calibration error.

5. The composite group number assembled at the University of Oulu by *Usoskin et al.* (2016), $R_{UEA}$. This series directly calibrates all data to the number of groups, $N_G$, defined by observers of the RGO for 1900-1976 from the photo-heliographic plates. Note that, like $R_{BB}$, it does not employ the 12.08 scaling factor that was used in the generation of $R_G$. This series is unique in that it avoids using either daisy chaining or regression techniques, and makes no assumptions about linearity or proportionality between different datasets. For the test interval presented here (1920-1976), $R_{UEA}$ and $N_G$ are identical. Note also that original group number by Hoyt and Schatten (1994, 1998) [$R_G$] is also of the same in form as $N_G$ over this interval (being 12.08$N_G$ for the interval analysed here) and so tests of $R_G$ are not performed here as they would give identical answers to $R_{UEA}$.

The tested data series are summarised in Table 1. Figure 1(a) shows the five tested data series. Some are group numbers whereas others are Wolf sunspot numbers and they employ different scaling factors, as discussed above: hence, so that they can be compared in figure 1, each has been regressed against the RGO sunspot group number, $N_G$ over the interval 1921–1945. The start date of this interval is chosen to be after any interval when there are some concerns over the calibration of the RGO data (Cliver and Ling, 2015), the end date is just before the Waldmeier discontinuity (Svalgaard, 2011; Clette and Lefèvre, 2016). Figure 1(a) shows that before 1946 (the vertical dot-dash line) all the series are either identical (other than the scaling factors) or very similar indeed. In fact, $R_C$ is, by its definition, identical to $R_{ISNv1}$ between 1848 and 1945 the scaled $R_{ISNv2}$ is found to also be virtually identical to $R_{ISNv1}$ for the interval 1921-1946. After 1946 it can be seen that these scaled variations diverge. Because some of the differences are rather small in figure 1(a), figure 1(b) shows the deviations of each from the mean of the five scaled sequences, $\Delta[R_G]_{fit}$. The Waldmeier discontinuity is clear in $R_{ISNv1}$ because after 1946 there are large positive values of this deviation around each sunspot maxima. Both $R_{ISNv2}$ and $R_{BB}$ show similar variations, but of the opposite sense to those for $R_{ISNv1}$; the variations for $R_{BB}$ being larger than those for $R_{ISNv2}$. These deviations for $R_C$ and $R_{UEA}$ oscillate around zero.

Figure 2 analyses the regressions over 1921-1945 used to put all variations on the same scale in figure 1(a). The best-fit regression lines in the scatter plots shown in the left-hand panels of this plot do not pass through the origin, showing linear, but not proportional dependencies between the tested sunspot numbers and the RGO group number, $N_G$. Note that $R_{ISNv1}$ and $R_C$ are identical over this regression interval and so both are analysed in the top row of figure 2. Note also that no plot is given for $R_{UEA}$ as it equals $N_G$ for this regression interval.

Great care must be taken when using linear regressions. For example, errors caused by inadequate and/or inappropriate regression techniques were discussed by Lockwood et al. (2006) in relation to differences between reconstructions of the magnetic field in near-Earth space from geomagnetic activity data. Nau (2016) has neatly summarised the problems: "If any of the assumptions is violated (i.e., if there are nonlinear relationships between dependent and independent variables or the errors exhibit correlation, heteroscedasticity, or non-normality), then the forecasts, confidence intervals, and scientific insights yielded by a regression model may be (at best) inefficient or (at worst) seriously biased or misleading." In the context of sunspot numbers and sunspot group numbers, Lockwood et al (2016c) found that the biggest problems were associated with non-normal distributions of data errors (especially if linearity or proportionality was inappropriately assumed) which violate the assumptions made by most regression techniques: such errors should always be tested for before a correlation is used for any scientific inference or prediction (Lockwood et al., 2006, 2016c). A normal distribution of fit residuals can be readily tested for using a Quantile–Quantile ("Q–Q") plot (e.g. Wilk and Gnanadesikan, 1968). This is a graphical technique for determining if two datasets come from populations with a common distribution; hence by making one of the datasets normally distributed we can test the other to see if it also has a normal distribution. The left hand panel of figure 2 gives the corresponding quantile-quantile (Q-Q) plots in which the ordered standardised fit residuals, $e_{(i|n)}/\sigma$ (where $\sigma$ is their standard deviation), are plotted as a function of quantiles of standard normal distribution, $F_N^{-1}[i-0.5/n]$. To be a reliable and useable regression fit, the points in a Q-Q plot should form a straight line along the diagonal as this shows the errors in the fitted data form a Gaussian distribution which is one of the assumptions of least squares regression fitting. It can be seen that this condition is reasonably well met for $R_{ISNv1}$ (and hence $R_C$) and $R_{ISNv2}$ (which are actually almost identical in form over the interval used) but not for $R_{BB}$ (figure 2d). Hence the error distribution for $R_{BB}$ is not Gaussian. The form of figure 2(d) suggests that the $R_{BB}$ distribution has a different kurtosis (sharpness of peak) compared to $N_G$ and is asymmetric. This actually applies for all of the $R_{BB}$ data series, but figure 2 shows that it even applies for the interval of the regression shown here (1921–1945), over which $R_{BB}$ and the other data series appear, at least visually, to be very similar (see Figure 1a). Hence figure 2 stresses that although some linear regressions give valid Q-Q plots, others do not. Hence, in general, linear regression fits cannot be relied upon and are used here in figure 1 for illustrative purposes only in displaying the tested data series.

*2.ii. Test Data Series*

We use five independent data series to test the various sunspot number sequences that are summarised in Table 2 and compared in figure 3.

1). The total spot area (corrected for limb foreshortening), $A_G$, from the RGO dataset (also called the Greenwich Photoheliographic Results, GPR) (Baumann and Solanki, 2005; Willis et al., 2013a; b). This dataset was compiled using white-light photographs (photoheliograms) of the Sun from a small network of observatories to produce a dataset of daily observations between 17$^{th}$ April 1874 and the end of 1976, thereby covering nine solar cycles. The observatories used were: The Royal Observatory, Greenwich (until 2$^{nd}$ May 1949); the Royal Greenwich Observatory, Herstmonceux (3$^{rd}$ May 1949 – 21$^{st}$ December 1976); the Royal Observatory at the Cape of Good Hope, South Africa; the Dehra Dun Observatory, in the North-West Provinces (Uttar Pradesh) of India; the Kodaikanal Observatory, in southern India (Tamil Nadu); and the Royal Alfred Observatory in Mauritius. Any remaining datagaps were filled using photographs from many other solar observatories, including the Mount Wilson Observatory, the Harvard College Observatory, Melbourne Observatory, and the US Naval Observatory. The sunspot areas were measured from the photographs with the aid of a large position micrometer (see Willis et al., 2013 and references therein). The $A_G$ values are the total sunspot area (umbrae plus penumbrae) and have been corrected for the effect of foreshortening which increases as sunspots are closer to the limb of the solar disc.

2). The number of groups, $N_G$, from the same RGO photographs as used to generate $A_G$. The RGO data did not employ the Zürich group classification scheme so $N_G$ is not influenced by the Waldmeier discontinuity. It is well known that the RGO group numbers show a drift relative to the Zürich sunspot numbers (e.g., Jakimcowa, 1966). This is not necessarily a calibration error as there are a number of ways in which it could have arisen from real changes in solar activity. The most obvious is that there has been a drift in the ratio of the number of individual spots to the number of spot groups which would influence $N_G$ and sunspot numbers differently. However, in addition, over the same interval there has been a drift in the lifetimes of spot groups, giving an increase in the number of recurrent groups (groups that are sufficiently long-lived to be seen for two or more traversals of the solar disc as seen from Earth) (Henwood et al., 2010). This has the potential to have influenced group numbers derived using different classification schemes in different ways.

3). The Mount Wilson CaK index. Spectroheliograms in the ionized calcium K line Ca II K (393.37 nm) were obtained between 1915 and 1985 using the 60-foot solar tower at Mount Wilson Observatory as part of their solar monitoring programme. Calibration of these images is, however, not straightforward. A new and homogeneous index quantifying the area of plages and active network in the Ca II K line has been derived from the digitization of almost 40000 photographic solar images by Bertello et al. (2010) (here referred to as the CaK index, CaKi). Although these data are available up to 1985, there were changes to the calibration procedure employed with "step wedge exposures" used from 9$^{th}$ October 1961. Because we want no possibility of effects by inhomogenities in the data caused by such changes, and because for the purposes of this paper the later data are not required, we here only employ CaKi data from before this date. Note that the CaK index has a pronounced non-linear variation with sunspot numbers (e.g., Foukal et al., 2009).

4). Ionospheric F2 region critical frequencies observed at Slough, foF2. As discussed in Article 1 (Lockwood et al., 2016a), the location of Slough means that the variation over each year is dominated by the plasma loss rate (and so by thermospheric composition) giving a dominant annual variation, as opposed to the semi-annual variation that dominates at some other stations (Scott and Stamper, 2016), and a close variation with sunspot numbers. Additional effects, quantified by the area of white light faculae, are small for the Slough data (Smith and King, 1981) and Article1 shows the main effect of including them in quantifying the Waldmeier discontinuity is to increase noise levels. Hence in this paper, Slough foF2 values are used without allowance for facular areas. In Article 1 (Lockwood et al., 2106a) nine dayside Universal Times (UTs) were identified for which the correlation of foF2 with sunspot numbers (after the Waldmeier discontinuity) exceeds 0.99 for all sunspot data series tested. Rather than treat these as independent data series, we average the nine together in the present paper.

5). The Mount Wilson Observatory (MWO) sunspot group number, $N_{MWO}$. These have been compiled routinely from January 1917 onwards using the 150-foot solar tower telescope from sketches of the solar disc. These data did not use the Zürich group classification scheme, employing instead the scheme originally developed by Hale and co-workers (Hale et al., 1919). Thus $N_{MWO}$ will not be influenced by the Waldmeier discontinuity. Because of different equipment and procedures, $N_{MWO}$ does not vary linearly with $N_G$.

The test data series are shown in figure 3 in the same format as figure 1 and are summarised in Table 2. In figure 3(a) each variation has been scaled to the variation of $N_G$ using regressions over the interval 1932-1961 (except foF2 for which data availability makes this interval 1933-1961). Because some proxies for solar activity, such as the CaK index, do not vary linearly with sunspot numbers, the fits are made using a 2nd-order polynomial. The coefficients for the derived second-order polynomial fits are given in Table 2. The analysis presented in tis paper was repeated for third-order polynomial fits and the results were essentially identical. The deviations from the mean of the five are shown in figure 3(b). Deviations are comparable to those in figure 1 before 1945 (but are considerably smaller than them for after 1945). They are also random in nature in that they are generally largest in single years and of the same general character before 1945 as after. Differences are largest at the start of the interval shown in figure 3. Figure 4 shows the regression scatter plots of these best polynomial fits and the corresponding Q-Q plots. It can be seen that the fit residuals for CaKi and foF2 are slightly non-Gaussian in the tails but the bulk of the population follows a normal distribution. There is a slight deviation from a normal distribution of errors for $A_G$, but $N_{MWO}$ gives an almost perfect normal distribution of errors. Hence all the test series show a near Gaussian distribution of errors when compared to $N_G$. The scatter plots show that the polynomial fits remove non-linearity and further tests on the residuals (see Lockwood et al., 2006) reveal they show neither correlation nor heteroscedasticity. Because they pass all these tests, the regressions between the fitted test series can be safely employed. In particular, the Q-Q plots shown in figures 2 and 4 justify the use the parametric (i.e., assuming a Gaussian distribution) t-test on the fit residuals described in the next section, (and explains why non-parametric tests give very similar answers).

*2.iii The analysis procedure*

Article 3 (Lockwood et al., 2016c) shows that it is important not to force regression fits between different sunspot number sequences through the origin of the scatter plot. Doing so means that proportionality (and hence linearity) between the sequences is assumed and results in the inflation of solar cycle amplitudes in data from a lower-acuity observer. Furthermore, Lockwood et al. (2016c) and Usokin et al. (2016) show that results different observers often have a non-linear dependence. Most previous studies of the Waldmeier discontinuity (Svalgaard, 2011; Lockwood et al., 2014a; Lockwood et al., 2016a; Clette and Lefèvre, 2016) implicitly made the assumption of proportionality because they assumed that correction for the Waldmeier discontinuity could be achieved using a single multiplicative factor. In this paper, we do not make this assumption, but evaluate a correction for before the Waldmeier discontinuity from $R$ to $R'$ which is given by equation (1). Adjusting the values before the putative Waldmeier discontinuity with the optimum $f_R$, $n$ and $\delta$ means that the sequence of older data is made consistent with the post-discontinuity data.

Clette and Lefèvre (2016) make the valuable point that the precise date of the Waldmeier discontinuity is not known and this can influence the results if the "before" and "after" intervals used in the method of Lockwood et al. (2014a) end and start, respectively, at an assumed date for the discontinuity. (This is because if that date is wrong, some data that is actually from before the discontinuity can be placed in the after interval, or vice-versa). Here we remove this dependency by ending the "before" interval in 1943 and starting the "after" interval in 1949. Thus, the precise date or waveform of discontinuity does not have an effect, provided the bulk of it is within the 6-year interval around 1946, which is the most likely date defined by Clette and Lefèvre (2016). The length of the "before" and "after" intervals was varied until an optimum was achieved, as discussed below.

The procedure used was to first determine the exponent $n$ and offset $\delta$ required by equation (1). Because these relate to the correction needed for a given tested sunspot number series, the same values of $n$ and $\delta$ are used when testing against all five test series. These values were obtained using the Nelder-Mead search procedure to find the optimum combination of $n$, $\delta$ and $f_R$ that made $R'$ correlate best with each of the test data series for between the start of the "before" interval and the end of the "after" interval. Because the test series are so similar (see figure 3), they gave very similar optimum $n$, $\delta$ and $f_R$ values and the $n$ and $\delta$ values adopted here were those for the test series that gave the highest correlation (which was invariably for the RGO sunspot group number, $N_G$). Having defined the optimum $n$ and $\delta$ values, the procedure used was to vary the factor $f_R$ between 0.5 and 1.3 (in steps of 0.001) to evaluate the mean fit residuals in the "before" and "after" intervals.

As in Lockwood et al (2014a), Welch's t-test is used to evaluate the probability $p$-values of the difference between the mean fit residuals (between the tested and test series in question) for the "before" and "after" intervals being zero. This two-sample *t*-test is a parametric test (i.e., it assumes a Gaussian distribution) that compares two independent data samples (Welch, 1947).

Because it is not assumed that the two data samples are from populations with equal variances, the test statistic under the null hypothesis has an approximate Student's t distribution with a number of degrees of freedom given by Satterthwaite's approximation (Satterthwaite, 1946). The distributions of residuals were found to be close to Gaussian in most cases and so application of non-parametric tests (specifically, the Mann-Whitney U (Wilcoxon) test of the medians and the Kolmogorov-Smirnov test of the overall distributions) gave very similar results. The overall pdf, $p(f_R)$, for the five test data series combined, is obtained by taking the product of those for each individually:

$$p_o(f_R) = [p(f_R)]_{N_G} \times [p(f_R)]_{A_G} \times [p(f_R)]_{CaK} \times [p(f_R)]_{N_{MWO}} \qquad (2)$$

The peak value of $p(f_R)$, $p_m$, is then defined. Note that $[p(f_R)]_{foF2}$ was not used for reasons discussed later. (However, including a term in $[p(f_R)]_{foF2}$ makes very little difference to the results).

Another valuable point made by Clette and Lefèvre (2016) is that if the "before" and "after" intervals are too long in duration then other errors (such as the Locarno calibration error in the case of $R_{ISNv1}$) can enter into both the tested and test series and so influence the estimate of the discontinuity correction. On the other hand, if these intervals are too short, the inter-annual variability due to "geophysical noise" in both the test and tested data will also degrade the final value. Hence an optimum compromise is needed. To reduce the number of variables, the "before" and "after" intervals were assigned the same duration, $T$. The value of $T$ was then varied between 1 yr and 23 yrs (the latter using all the test data shown in figure 3, except for the 6-year interval around the putative Waldmeier discontinuity). As expected from the above, both the lowest and the largest $T$ values gave a low peak values of $p_m$, and hence broad distributions of $p_o(f_R)$. The narrowest $p_o(f_R)$ distribution, giving the largest peak value $p_m$, was for $T = 11$ years (approximately one full solar cycle). Hence we use a "before" interval of 1932–1943 and an "after" interval of 1949–1960, as this minimised the width of the overall probability distribution function obtained, and hence the uncertainties. This is the optimum compromise between having sufficient data points and minimising the potential to introduce other errors and discontinuities present in either data series.

The second, subsidiary, test used by Lockwood et al. (2014a; 2016) employed the correlations, $r$, between $R'$ and each of the test series over the whole interval (1932-1960). The peak in $r$ will be when the discontinuity introduced into $R'$ most closely cancels that inherent in the data series $R$: values of $r$ will be lower for less-than-optimum combinations of $f_R$, $n$ and $\delta$. The peaks of the correlograms ($r$ against $f_R$ for the optimum $\delta$ and $n$) were defined and for each $f_R$ the significance $S$ of the difference between $r$ and its peak value was quantified using the Fischer-Z transform by comparison against the AR-1 noise model. These significance values are then combined into an overall variation for all five test series by multiplying the probabilities:

$$S_o(f_R) = 1 - \{ (1-[S(f_R)]_{N_G}) \times (1-[S(f_R)]_{A_G}) \times (1-[S(f_R)]_{CaK}) \times (1-[S(f_R)]_{N_{MWO}}) \} \qquad (3)$$

Note that, as for $p_o(f_R)$, the term $[S(f_R)]_{foF2}$ has been omitted in equation (3) (but, again, its inclusion makes very little difference). Ideally, the minimum in $S_o(f_R)$ would be at the same $f_R$ as the peak in $p(f_R)$. A minimum $S_o(f_R)$ of zero would indicate perfect agreement between the results of this second test for all five test data series.

## 3. Results

### 3.i Results for $R_{ISNv1}$

Figure 5 summarises the results of these tests for version 1 of the SIDC Wolf/Zurich/International sunspot number composite (i.e., $R$ in equation (1) is $R_{ISNv1}$). The figure is for the optimum value of $\delta$ and $n$ which are found to be 2.731 and 1.088, respectively. The various coloured lines in the top panel show the correlation coefficients between the adjusted $R_{ISNv1}$ series, $R_{ISNv1}'$, and the test series for annual means over the full internal (1932-1960), $r$, as a function of $f_R$ for this $\delta$ and $n$. The best correlation is for the number of spot groups from the RGO data, $N_G$ (in mauve). Peak correlation occurs at the same $f_R$ for the polynomial-fitted CaK index $[CaKi]_{fit}$ (in blue) and the fitted Mount Wilson sunspot group number, $[N_{MWO}]_{fit}$ (in red). The peak for the fitted total spot area from the RGO data, $[A_G]_{fit}$ (in green) is at a slightly lower $f_R$ and for the fitted average Slough F2 layer critical frequency $[foF2]_{fit}$ at a yet lower $f_R$ (in orange).

The middle panel of figure 5 shows the statistical significances of the difference between the $r$ at general $f_R$ and the peak value using the same colour scheme. The black line shows the overall significance $S_o(f_R)$, given by equation (3).

The bottom panel of figure 5 shows the *p*-values of the differences in fit residuals between for the "before" and "after" intervals for the fits of the adjusted tested series $R'$ and each test series, again using the same colour scheme. The black line shows the overall pdf $p_o(f_R)$, given by equation (2). It can be seen that the minimum in the combined $S_o(f_R)$ and the peak in the combined $p_o(f_R)$ are at very similar $f_R$ and so the two tests are in excellent agreement. The uncertainty in the optimum value for $S_o(f_R)$ is much greater than that for $p_o(f_R)$ (the distribution being much broader) and so $p_o(f_R)$ provides the most stringent test for the optimum $f_R$ value. The grey band marks the 2-sigma points of the $p_o(f_R)$ distribution. Note that the agreement of the $f_R$ of minimum $S_o(f_R)$ and peak $p_o(f_R)$ is less close for $[foF2]_{fit}$ (in orange). Thus the foF2 test series is the only one for which the two tests do not completely agree. This was found to be true for all of the tested sunspot series. For this reason foF2 is left out of the computation of both $p_o(f_R)$ and $S_o(f_R)$ (equations 2 and 3). The orange lines in figure 4 (and subsequent figures) do, however, serve to show that this terrestrial proxy for solar activity gives results that are still (just) within the $2\sigma$ uncertainty band derived from the four more direct solar indices. Indeed, all the test data series give results within the $\pm 2\sigma$ uncertainty. The optimum combination of $f_R$, $n$ and $\delta$ defined by figure 5 is given for this tested series ($R_{ISNv1}$) in the top row of Table 3.

The optimum correction for the Waldmeier discontinuity for $R_{ISNv1}$ is:

$R_{ISNv1}' = R_{ISNv1}$ for $t \geq 1946.0$

$R_{ISNv1}' = 0.7350 \times R_{ISNv1}^{1.0883} + 2.7308$ for $t < 1946.0$  (4)

Applying equation (4) gives a mean $\langle R_{ISNv1}'\rangle$ of 61.23 over the "before" interval, whereas $\langle R_{ISNv1}\rangle$ is 54.53 over the same interval. Hence this test shows that $R_{ISNv1}$ is 12.28% too low in the "before" interval. This, like previous studies, confirms that the Waldmeier discontinuity is a real factor in $R_{ISNv1}'$. Using the 2σ points for $f_R$ yields an uncertainty in the 12.28% error of ±3.37%. The percent change is only slightly greatly than the 11.9% correction found in the studies by Lockwood et al. (2014, 2016a), despite the several improvements and refinements to the method that have been made in the present. The optimum value is smaller than the 15.8% derived by Clette and Lefèvre (2016) for $R_{ISNv1}$ which is close to, but just outside, the upper edge of the 2σ uncertainty band found here. As in previous studies by Lockwood et al. (2014, 2016a), the probability that the required change is the 20% originally invoked by Svalgaard (2011) is essentially zero. However, notice that the neither the zero-level offset nor the exponent is small: hence the Waldmeier discontinuity in $R_{ISNv1}$ requires a non-linear corrections and a proportional (i.e., multiplicative) one is not adequate.

### 3.ii Results for $R_{ISNv2}$

Figure 6 is the equivalent plot to figure 5 for the new version of the SIDC Wolf/Zurich/International sunspot number composite, $R_{ISNv2}$. The behaviour is very similar to figure 5 other than the peak of the pdf is at $f_R = 0.9967$ for δ = 0.0001 and $n = 0.9967$ (see Table 3). Again, the level of agreement between the results for the different test series is exceptionally good. The optimum correction is:

$R_{ISNv2}' = R_{ISNv2}$ for $t \geq 1946.0$

$R_{ISNv2}' = 0.9760 \times R_{ISNv2}^{0.9967} + 0.0001$ for $t < 1946.0$  (5)

This test finds that $R_{ISNv2}$ overestimates the mean for the "before" interval by 3.80±2.91%. Thus the Waldmeier discontinuity has been slightly overestimated in $R_{ISNv2}$. Note that ideal value of zero is (just) outside the 2σ uncertainty for $R_{ISNv2}$. The very small δ and the closeness of $n$ to unity mean that the correction needed is very close to being proportional. Hence the correction in $R_{ISNv2}$, although slightly too large, has removed the non-linearity introduced by the changes made by Waldmeier.

### 3.iii Results for $R_C$

Figure 7 is the equivalent plot to figure 5 for the corrected Wolf/Zurich/International sunspot number composite proposed by *Lockwood et al.* [2015]. The peak of the pdf is at $f_R = 0.6240$ for δ = 3.4957 and $n = 1.0950$ (see Table 3). Again, the level of agreement between the results for the different test series is exceptionally good. The optimum correction is:

$R_C' = R_C$ for $t \geq 1946.0$

$$R_C' = 0.6240 \times R_C^{1.0950} + 3.4957 \text{ for } t < 1946.0 \qquad (6)$$

This test finds that $R_C$ underestimates the mean for the "before" interval by 0.44±3.01%. Although this underestimate is zero to within the 2σ uncertainty, the correction for the Waldmeier discontinuity in $R_C$ is nevertheless less satisfactory than that in $R_{ISNv2}$. This is because, as for $R_{ISNv1}$, the value of $n$ is not close to unity and so the non-linear behaviour introduced by the Waldmeier discontinuity has not been removed.

### 3.iv Results for $R_{BB}$

Figure 8 is the equivalent plot to figure 5 for the new backbone sunspot group number composite proposed by Svalgaard and Schatten (2015), $R_{BB}$. The peak of the pdf is at $f_R = 0.7410$ for $\delta = 0.3108$ and $n = 1.0932$ (see Table 3). Again, the level of agreement between the results for the different test series is exceptionally good. The optimum correction is:

$$R_{BB}' = R_{BB} \text{ for } t \geq 1946.0$$

$$R_{BB}' = 0.7410 \times R_{BB}^{1.0932} + 0.3108 \text{ for } t < 1946.0 \qquad (7)$$

This test finds that $R_{BB}$ overestimates the mean for the "before" interval by 5.74±2.25%. In addition to this being significantly different from zero, the correction for the Waldmeier discontinuity is, as for $R_C$, less satisfactory than that in $R_{ISNv2}$ because the value of $n$ is not as close to unity and so any non-linear behaviour introduced by the Waldmeier discontinuity has not been removed.

### 3.iv Results for $R_{UEA}$

Figure 9 is the equivalent plot to figure 5 for the new Usoskin et al. (2015) group number reconstruction, $R_{UEA}$. We would expect this to give $f_R$ very close to unity and δ very close to zero because, for the interval studied in his paper, $R_{UEA} = N_G$ and hence is the same as one of the four test data sequences. However, the test is interesting as it shows the net effect of the other solar test sequences (the CaK index, the RGO spot areas $A_G$, and the Mount Wilson group numbers $N_{MWO}$) on the result is negligible and also shows the same behaviour for foF2 as the other tested sunspot series. The top panel of figure 9 shows the unity peak $r$ between $R_{UEA}$ and $N_G$, but other than this the behaviour for the other test series is very similar to that for the other tested series. In this case, the $p_o(f_R)$ curve is essentially a delta function (the plot scale in figure 9(c) is the same as for parts (c) of figures 5-8, but the peak value of $p_o(f_R)$, $p_m$, is off-scale in his case as it is close to unity). To within 4 decimal places, values of $n$ and $f_R$ are unity and δ is zero. The change required in the "before" interval is 0.005±0.048%.

This test of $R_{UEA}$ shows that the procedure works well and that when presented with one dominant correlation, the other test series, that give slightly different optimum $f_R$, do not degrade the result.

### 3. Conclusions

We have tested five sunspot data series around the putative Waldmeier discontinuity in sunspot numbers around 1945 using 5 diverse test datasets that are all completely independent of the Zürich sunspot number which are the source of this discontinuity. The test data are: the sunspot group number from the RGO dataset $N_G$, the total sunspot area from the RGO dataset (corrected for foreshortening) $A_G$, the Mount Wilson CaK index, the Mount Wilson sunspot group number $N_{MWO}$, and the ionospheric F2 region critical frequency observed at Slough foF2. We have tested various sunspot data series in two ways, using the fit residuals and using the correlation coefficient. In all cases, the results of these two methods are remarkably consistent but the uncertainties are lower for the fit residual method. The most persistent difference between the two methods occurs for the ionospheric foF2 data which is here not included in overall tests but are nevertheless plotted to show that these terrestrial data still give results that are consistent with those for the solar test data to within the 2σ uncertainties. The diversity of the derivations and sources of these test series means that the chances that all suffer from the same error around 1946 are negligible and comparison shows random data noise differences between them (figure 3) and not systematic errors.

To summarise our results graphically, figure 10 plots the variations of all the tested series over the test period (which covers solar cycles 17, 18 and 19). The grey area is the mean of the 5 test series, and in parts (b)-(f) the blue line is the tested series and the red line is the tested series with the relevant adjustment to the data before 1946, as derived in this paper. Part (a) of figure 10 shows the five test series.

Figure 10(b) is for $R_{ISNv1}$ and the Waldmeier discontinuity is clearly visible in the blue line as low values during cycle 17. The red line demonstrates how effective the correction is – and this is true for all the tested series. Figure 10(c) is for $R_{BB}$ and the blue line shows that values in cycle 17 are persistently too large. It is not at all clear how this has occurred because $R_{BB}$ was compiled from various observers, most of whom did not change practices in defining groups when such changes were made at Zürich. However it appears that $R_{BB}$ has somehow been adjusted to allow for the Waldmeier discontinuity and that adjustment is either not warranted or excessive. Figure 10(d) is for $R_{ISNv2}$ and the Waldmeier discontinuity is much reduced compared to $R_{ISNv1}$. However there appears to be a slight over-correction for the discontinuity as values for cycle 17 are slightly too large. This is consistent with the estimated inflation factors used to correct $R_{ISNv1}$ which was 18% (Clette and Lefèvre, 2016) which is larger than the value for the mean of $R_{ISNv1}$ over cycle 17 of 12.28± 3.37 that was derived here. Figure 10(d) confirms the effects of the mean for $R_{ISNv2}$ for cycle 17 being too large by the 3.80±2.91% that was derived in this paper. Figure 10(e) shows the results for $R_C$ and, although a good match to the mean for cycle 17 is obtained, the effects of the residual non-linearity can be seen with values at both sunspot minimum and sunspot maximum being a little low in $R_C$. Figure 10(f) shows the effects of the mean for $R_{UEA}$ – because the tested series and one of the test series are the same here the blue and red lines are essentially identical and both match the man test series very well.

Table 3 gives the optimum corrections needed for the five tested sunspot data series. Direct and careful allowance for this discontinuity has been made in version of the Wolf/Zurich/International sunspot number, $R_{ISNv2}$, but we here show that the correction applied is slightly too large but does remove the non-linearity inherent in $R_{ISNv1}$. Note that because $R_{ISNv2}$ is compiled by daisy-chaining of calibrations, this systematic error will be passed to all prior data. The correction used in the "backbone" sunspot group series, $R_{BB}$, of Svalgaard and Schatten (2016) is too also large. A large part of this is likely to be the 7% correction introduced by Svalgaard and Schatten to allow for the "evolutionary" aspect of Waldmeier's classification scheme but it is not at all obvious that this is required for the data used to compile $R_{BB}$. The backbone series is the only one not to give a usable Q-Q plots when regressed against other sunspot series. From the analysis presented in Paper 3 (Lockwood et al., 2015c) we think that some of the error has arisen from the use of linear inter-correlation of segments of annual mean data (when in general the relationship is non-linear) and because fits were unnecessarily forced through the origin which tends to amplify solar cycle amplitudes in fitted data. As for $R_{ISNv2}$, $R_{BB}$ uses daisy-chaining of calibrations and this error will be passed to prior data and such errors will accumulate as one goes back in time.

The correction applied by Lockwood et al. (2014a) to $R_{ISNv1}$ to generate $R_C$ is designed to remove the Waldmeier discontinuity on average data series. These tests show that this is achieved, but that the non-linear variation with the test data, as also found for $R_{ISNv1}$ has not been removed. In addition, $R_C$ only considered two known errors and others certainly exist, for example the modern values were not corrected for the drift in the Locarno calibration values (Clette et al., 2015).

**Acknowledgements**   The authors are grateful to staff and funders of the World Data Centres from where data were downloaded: specifically, the Slough foF2 data were obtained from WDC for Solar Terrestrial Physics, part of the UK Space Science Data Centre (UKSSDC) at RAL Space, Chilton, and the $R_{ISNv1}$ and $R$ISNv2, data from the WDC for the sunspot index, part of the Solar Influences Data Analysis Center (SIDC) at Royal Observatory of Belgium. We also thank David Hathaway and the staff of the Solar Physics Group at NASA's Marshall Space Flight Center for maintaining the on-line database of RGO data used here. Other sunspot data used ($R_{BB}$, $R_C$, $N_{MWO}$ and $R_{UEA}$) were taken from the respective cited publications. This work has been funded by STFC consolidated grant number ST/M000885/1.

# References

Asvestari, E., Usoskin, I.G., Kovaltsov, G.A., Owens, M.J. and Krivova, N.A.: 2016, Validation of the sunspot (group) number series against the cosmogenic isotope records, *Solar Phys.*, submitted

Baumann, I., & Solanki, S.K.: 2005, On the size distribution of sunspot groups in the Greenwich sunspot record 1874-1976, *Astron. & Astrophys.,* 443 (3), 1061-1066, doi: 10.1051/0004-6361:20053415

Bertello, L., Ulrich, R.K., & Boyden, J.E.: 2010, The Mount Wilson Ca II K Plage Index Time Series, *Solar Phys.*, 264, 31–44, doi: 10.1007/s11207-010-9570-z

Çakmak, H.: 2014, A digital method to calculate the true areas of sunspot groups, *Exp. Astron.*, 37, 539–553, doi: 10.1007/s10686-014-9381-6

Clette, F. & Lefèvre, L.: 2016, The new Sunspot Number: assembling all corrections, Solar Phys., in press. arXiv:1510.06928

Clette, F., Berghmans, D., Vanlommel, P., Van der Linden, R.A.M., Koeckelenbergh, A., & Wauters, L.: 2007, From the Wolf number to the International Sunspot Index: 25 years of SIDC, *Adv. Space Res.*, 40, 919–928. doi:10.1016/j.asr.2006.12.045

Clette, F., Svalgaard, L., Vaquero, J.M., & Cliver, E.W.: 2015, Revisiting the sunspot number, *in "The Solar Activity Cycle"*, eds. A. Balogh, H. Hudson, K. Petrovay and R. von Steiger, 35, Springer, New York. doi: 10.1007/978-1-4939-2584-1_3

Cliver, E, & Ling, A.G.: 2016, The discontinuity circa 1885 in the group sunspot number. *Solar Phys.* in press, doi: 10.1007/s11207-015-0841-6

Foukal, P., Bertello, L., Livingston, W.C., Pevtsov, A.A., Singh, J., Tlatov, A.G., & Ulrich, R.K.: 2009, A Century of Solar Ca II Measurements and Their Implication for Solar UV Driving of Climate, *Solar Phys.*, 255: 229–238, doi: 10.1007/s11207-009-9330-0

Hale, G.E., Ellerman, F., Nicholson, S.B., & Joy, A.H.: 1919, The Magnetic Polarity of Sun-Spots, *Astrophys. J.,* 49, 153-178.

Henwood, R., Chapman, S.C., & Willis, D.M.: 2010, Increasing Lifetime of Recurrent Sunspot Groups Within the Greenwich Photoheliographic Results, *Solar Phys.*, **262** (2) 299-313, doi: 10.1007/s11207-009-9419-5

Hoyt, D.V., Schatten, K.H., & Nesme-Ribes, E.: 1994, The one hundredth year of Rudolf Wolf's death: Do we have the correct reconstruction of solar activity?, *Geophys. Res. Lett.,* 21(18), 2067–2070, doi:10.1029/94GL01698.

Hoyt, D.V., & Schatten, K.H.: 1998, Group sunspot numbers: A new solar activity reconstruction, *Solar Phys.*, 181(2), 491–512, doi:10.1023/A:1005056326158.

Jakimcowa, M.: 1966, An Analysis of Differences in Number of Sunspot Groups Between the Greenwich and Zürich Observations, *Acta Astronomica*, 16(4), 317-328

Kiepenheuer, K.O.: 1953, Solar Activity, *in "The Sun",* ed. Gerard P. Kuiper, The University of Chicago Press, 1953, 322

Lefèvre, L., & Clette, F.: 2014, Survey and Merging of Sunspot Catalogs, *Solar Phys*, 289, 545–561. doi: 10.1007/s11207-012-0184-5

Leussu, R., Usoskin, I.G. & Mursula K.: 2013, Inconsistency of the Wolf sunspot number series around 1848, *Astron.& Astrophys.*, 559, A28, doi:10.1051/0004-6361/201322373.

Lockwood, M., Owens, M.J., Barnard, L.: 2014a, Centennial variations in sunspot number, open solar flux, and streamer belt width: 1. Correction of the sunspot number record since 1874, *J. Geophys. Res. Space Physics*, **119**, 5193. doi:10.1002/2014JA019970.

Lockwood, M., Owens, M.J., Barnard, L.: 2014b, Centennial variations in sunspot number, open solar flux, and streamer belt width: 2. Comparison with the geomagnetic data, *J. Geophys. Res. Space Physics*, 119, doi:10.1002/2014JA019972

Lockwood, M., Rouillard, A., Finch, I., Stamper, R.: 2006, Comment on "The IDV index: Its derivation and use in inferring long-term variations of the interplanetary magnetic field strength" by Leif Svalgaard and Edward W. Cliver, *J. Geophys. Res.: Space Physics* **111**, A09109. doi:10.1029/2006JA011640.

Lockwood, M., Scott, C.J., Owens, M.J., Barnard, L.A., & Willis, D.M.: 2016a, Tests of sunspot number sequences: 1. Using ionosonde data, *Solar Phys.*, in press, doi: 10.1007/s11207-016-0855-8

Lockwood, M., Owens, M.J., Barnard, L.A., Scott, C.J., Usoskin, I.G., and Nevanlinna, H.: 2016b, Tests of sunspot number sequences: 2. Using geomagnetic and auroral data, *Solar Phys.*, submitted

Lockwood, M., Owens, M.J., Barnard, L.A., & Usoskin, I.G.: 2016c, Tests of sunspot number sequences: 3. Effects of regression procedures on the calibration of historic sunspot data, Sol. Phys., *Solar Phys.*, doi: 10.1007/s11207-015-0829-2

Nau, R.: 2016, Statistical forecasting: notes on regression and time series analysis, http://people.duke.edu/~rnau/411home.htm

Scott, C.J. & Stamper, R.: 2015, Global variation in the long-term seasonal changes observed in ionospheric F region data, *Ann. Geophys.,* 33, 449–455, doi:10.5194/angeo-33-449-2015

Satterthwaite, F.E.: 1946, An Approximate Distribution of Estimates of Variance Components, *Biometrics Bulletin*, 2, 110–114, doi:10.2307/3002019

Smith, P., & King, J.W.: 1981, Long-term relationships between sunspots, solar faculae and the ionosphere, *J. Atmos. Terr. Phys.*, 43 (10), 1057–1063. doi: 10.1016/0021-9169(81)90020-9

Svalgaard, L.: 2011, How well do we know the sunspot number?, *Proc. Int. Astron. Union*, 7, 27–33, doi:10.1017/S1743921312004590.

Svalgaard, L., Cagnotti, M., Cortesi, S.: 2016, The effect of Sunspot Weighting, *Solar Phys.* submitted. ArXiv 1507.01119.

Taylor, P.O.: 1991, Observing the Sun, chapter 4, Cambridge University Press, ISBN 0-521-40110-0.

Usoskin, I.G., Mursula, K., & Kovaltsov, G.A.: 2003, Reconstruction of monthly and yearly group sunspot numbers from sparse daily observations, *Solar Phys.*, 218, 295-305, doi: 10.1023/B:SOLA.0000013029.99907.97

Usoskin, I.G., Kovaltsov, G.A., Lockwood, M., Mursula, K., Owens, M.J., Solanki, S.K.: 2016, A new calibrated sunspot group series since 1749: statistics of active day fractions. *Solar Phys.*, in press, doi: 10.1007/s11207-015-0838-1

Waldmeier, M.: 1947, Publications of the Zürich observatory, 9, p.1

Welch, B. L.: 1947, The generalization of "Student's" problem when several different population variances are involved, *Biometrika*, 34 (1–2): 28–35. doi:10.1093/biomet/34.1-2.28

Wilk, M.B., & Gnanadesikan, R.: 1968, Probability plotting methods for the analysis of data, *Biometrika*, 55 (1): 1–17, doi:10.1093/biomet/55.1.1

Willis, D.M., Coffey, H.E., Henwood, R., Erwin, E.H., Hoyt, D.V., Wild, M.N., Denig, W.F.: 2013a, The Greenwich Photo-heliographic Results (1874 – 1976): Summary of the observations, applications, datasets, definitions and errors, *Solar Phys.*, 288, 117. doi: 10.1007/s11207-013-0311-y.

Willis, D.M., Henwood, R., Wild, M.N., Coffey, H.E., Denig, W.F., Erwin, E.H., Hoyt D.V.: 2013b, The Greenwich Photo-heliographic Results (1874 – 1976): Procedures for Checking and Correcting the Sunspot Digital Datasets, *Solar Phys.*, 288, 141. doi: 10.1007/s11207-013-0312-x

Willis, D.M., Wild, M.N., Warburton, J.S.: 2016, Re-examination of the Daily Number of Sunspot Groups for the Royal Observatory, Greenwich (1874 – 1885), *Solar Phys.*, in press

Wolf, R.: 1861, Vortrag uber die Sonne und ihre Flecken, Mitth. über die Sonnenflecken, 12. doi: 10.3931/e-rara-3058

**Table 1**. Sunspot data series tested in the present paper.

| symbol | name | Brief description | reference(s) |
|---|---|---|---|
| $R_{ISNv1}$ | Wolf-Zürich-International sunspot number, version 1 | Sunspot number composite compiled at Zürich observatory and then the Royal Observatory of Belgium. Used as the standard series until July 2015 | Clette et al. (2007) Waldmeier (1947) Wolf (1861) |
| $R_C$ | Corrected sunspot number | $R_{ISNv1}$ with simple corrections for discontinuities at 1945 and 1849 | Lockwood et al. (2014a) |
| $R_{ISNv2}$ | Wolf-Zürich-International sunspot number, version 2 | Sunspot number composite from the same data as used to generate $R_{ISNv1}$ with a number of corrections. Used as the standard series after July 2015 | Clette et al. (2014) |
| $R_{BB}$ | Backbone sunspot group number | Sunspot group number composite compiled from various observers using the "backbone" method | Svalgaard & Schatten (2016) |
| $R_{UEA}$ | Usoskin et al. sunspot group number | Sunspot group number composite compiled from various observers using the statistics of active day fractions. It equals the RGO group number, $N_G$, for the interval tested here. | Usoskin et al. (2016) |

**Table 2.** Test data series used in the present paper and the coefficients of the best-fit 2$^{nd}$ order polynomial fit of test series $x$ to $N_G$, $[N_G]_{fit} = ax^2 + bx + c$.

| symbol | brief description | reference(s) | units | 2$^{nd}$ order polynomial fit coefficients | | |
|---|---|---|---|---|---|---|
| | | | | $a$ | $b$ | $c$ |
| $N_G$ | The number of sunspot groups identified from photographic plates by RGO observers | Willis et al. (2013a;b) | annual mean of daily number | 0 | 1 | 0 |
| $A_G$ | Corrected (for limb foreshortening) total sunspot area identified from photographic plates by RGO observers | Willis et al. (2013a;b) | 10$^{-6}$ of a solar hemisphere | −4.8253 × 10$^{-7}$ | 5.6452 × 10$^{-3}$ | 0.4232 |
| $N_{MWO}$ | The number of sunspot groups identified from solar drawings by MWO observers | Lefèvre & Clette (2014) Hale et al. (1919) | number of distinct groups in 10 months | −5.1202 × 10$^{-4}$ | 0.2070 | −0.5617 |
| CaKi | The CaK line index from MWO observations | Bertello et al. (2010) | - | 5.5864 × 10$^{-7}$ | 2.5681 × 10$^{-2}$ | −7.0114 |
| foF2 | The mean dayside ionospheric F2-layer critical frequency from the Slough ionosonde | Lockwood et al. (2016a) Smith & King (1981) | MHz | 6.5004 × 10$^{-3}$ | 2.2589 | −10.9406 |

**Table 3**. Optimum values of the fitted values of δ, *n* and $f_R$ (see equation 1) for the 5 tested sunspot data series.

| symbol | δ | *n* | optimum $f_R$ | Percent change required to "before" interval |
|---|---|---|---|---|
| $R_{ISNv1}$ | 2.7309 | 1.0884 | 0.7350 ± 0.0231 | +12.2787 ± 3.3692 |
| $R_C$ | 3.4957 | 1.0950 | 0.6240 ± 0.0198 | +0.4396 ± 3.0098 |
| $R_{BB}$ | 0.3108 | 1.0932 | 0.7410 ± 0.0191 | −5.7380 ± 2.2532 |
| $R_{ISNv2}$ | 1.4938×10$^{-4}$ | 0.9967 | 0.9760 ± 0.0295 | −3.7960 ± 2.9081 |
| $R_{UEA}$ | 0.0000 | 1.0000 | 1.0000 ± 4.7568×10$^{-4}$ | +0.0050 ± 0.0476 |

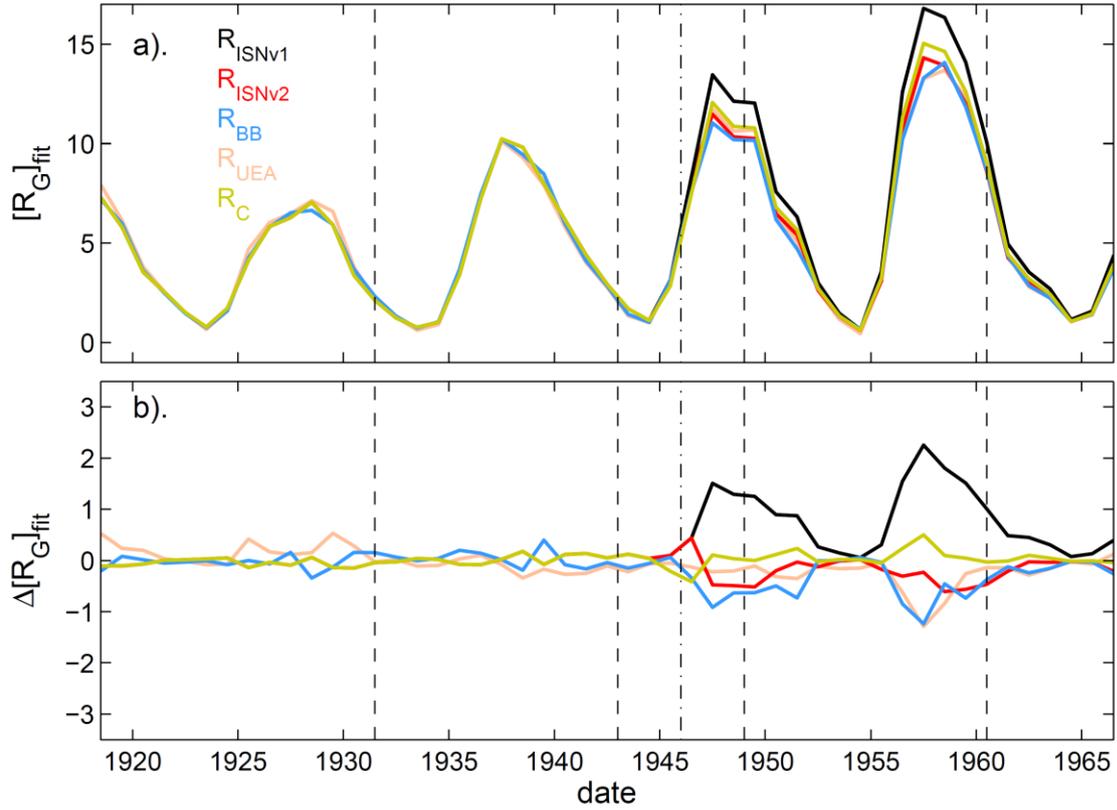

**Figure 1**. Comparison of the tested sunspot data series: (black) $R_{ISNv1}$; (red) $R_{ISNv2}$; (blue) $R_{BB}$; (pink) $R_{UEA}$; and (olive) $R_C$. To enable easy comparison all have been scaled by linear regression to the RGO sunspot group number $N_G$ over the interval 1921-1945. The top panel shows the regressed time series and the bottom panel shows he differences between each regressed variation and the average of the five scaled tested series. The vertical dot-dash line is the most likely time of the Waldmeier Discontinuity (1946) and the vertical dashed lines delineate the optimum "before" and "after" intervals found by the analysis.

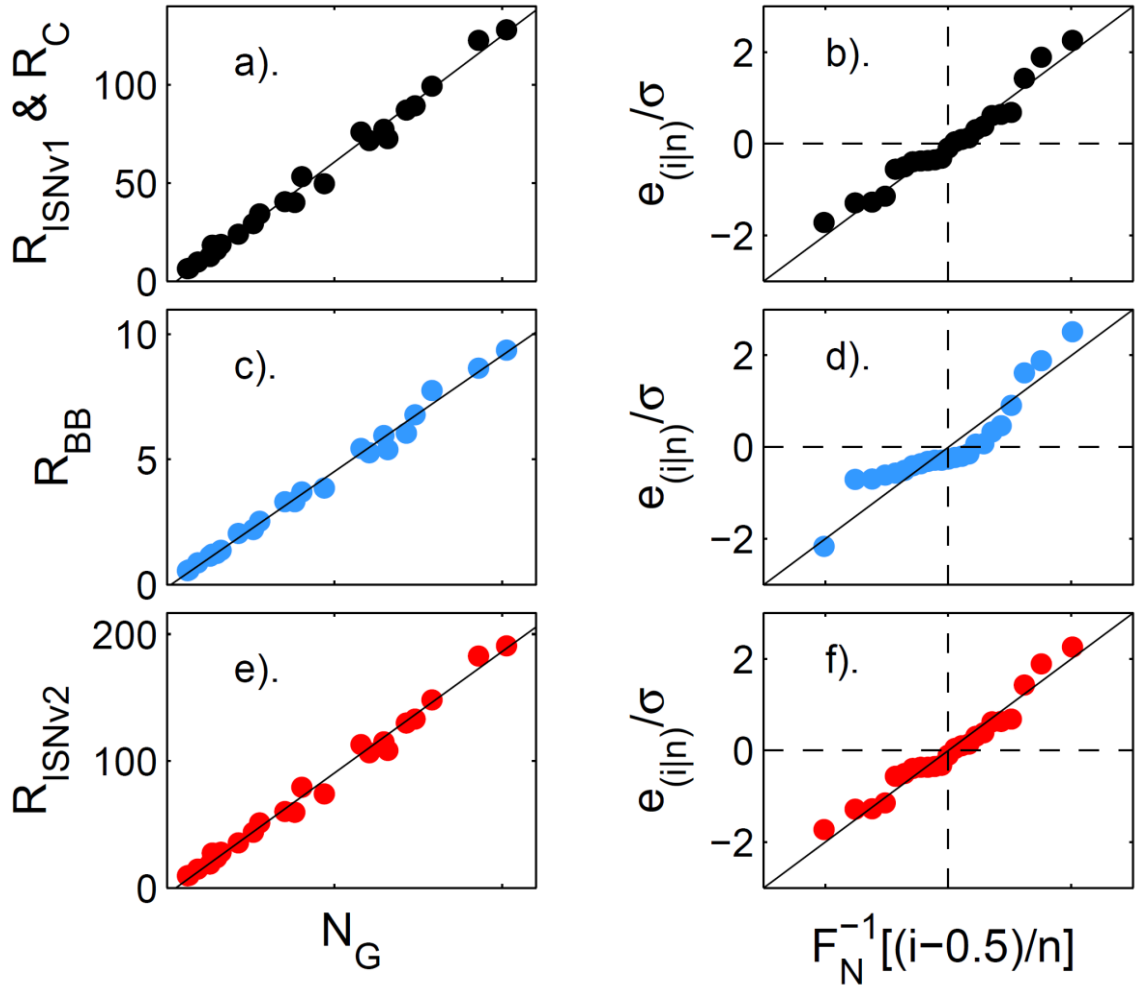

**Figure 2**. Analysis of the regressions between annual means $N_G$ and the independent tested sunspot data series over the interval 1921-1945, as used in figure 1. The left-hand panel (a, c, and e) show the scatter plots and the best-fit linear regression and the right hand panel (b, d, and f) the corresponding quantile-quantile (Q-Q) plots in which the ordered standardised fit residuals, $e_{(i|n)}/\sigma$, where $\sigma$ is their standard deviation) are plotted as a function of quantiles of standard normal distribution, $F_N^{-1}[i-0.5/n]$. (a) and (b) are for both $R_{ISNv1}$ and $R_C$ (which are identical over the regression interval); (c) an (d) are for $R_{BB}$; and (e) and (f) are for $R_{ISNv2}$. No plots are given for $R_{UEA}$ because over this interval it equals $N_G$.

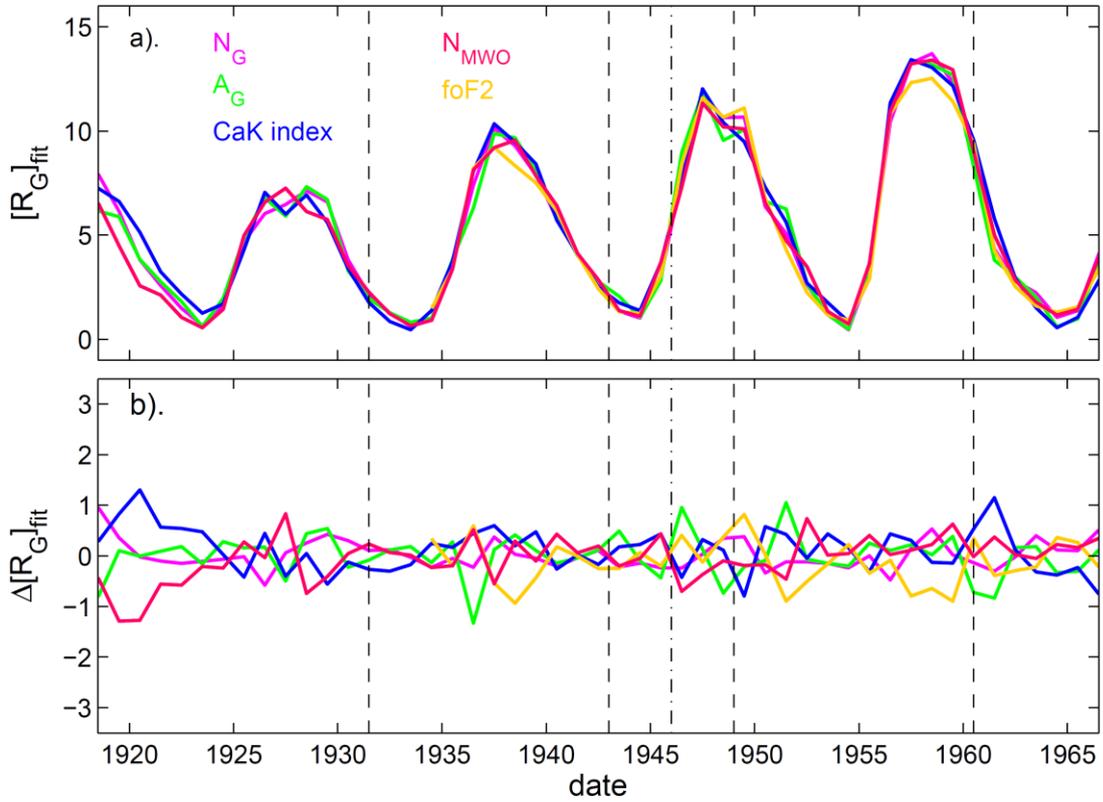

**Figure 3**. Comparison of the test data shown in the same format as figure 1: (mauve) the RGO group number, $N_G$; (green) the fitted RGO whole spot total area (corrected for foreshortening), $A_G$; (red) the fitted MWO group number, $N_{MWO}$; (blue) the fitted Mount Wilson CaK index; (orange) the fitted Slough F2 layer critical frequency, foF2. All series shown use a $2^{nd}$-order polynomial fit to the RGO $N_G$ data over 1921-1961. The vertical dot-dash line is the most likely time of the Waldmeier Discontinuity (1946) and the vertical dashed lines delineate the optimum "before" and "after" intervals found by the analysis.

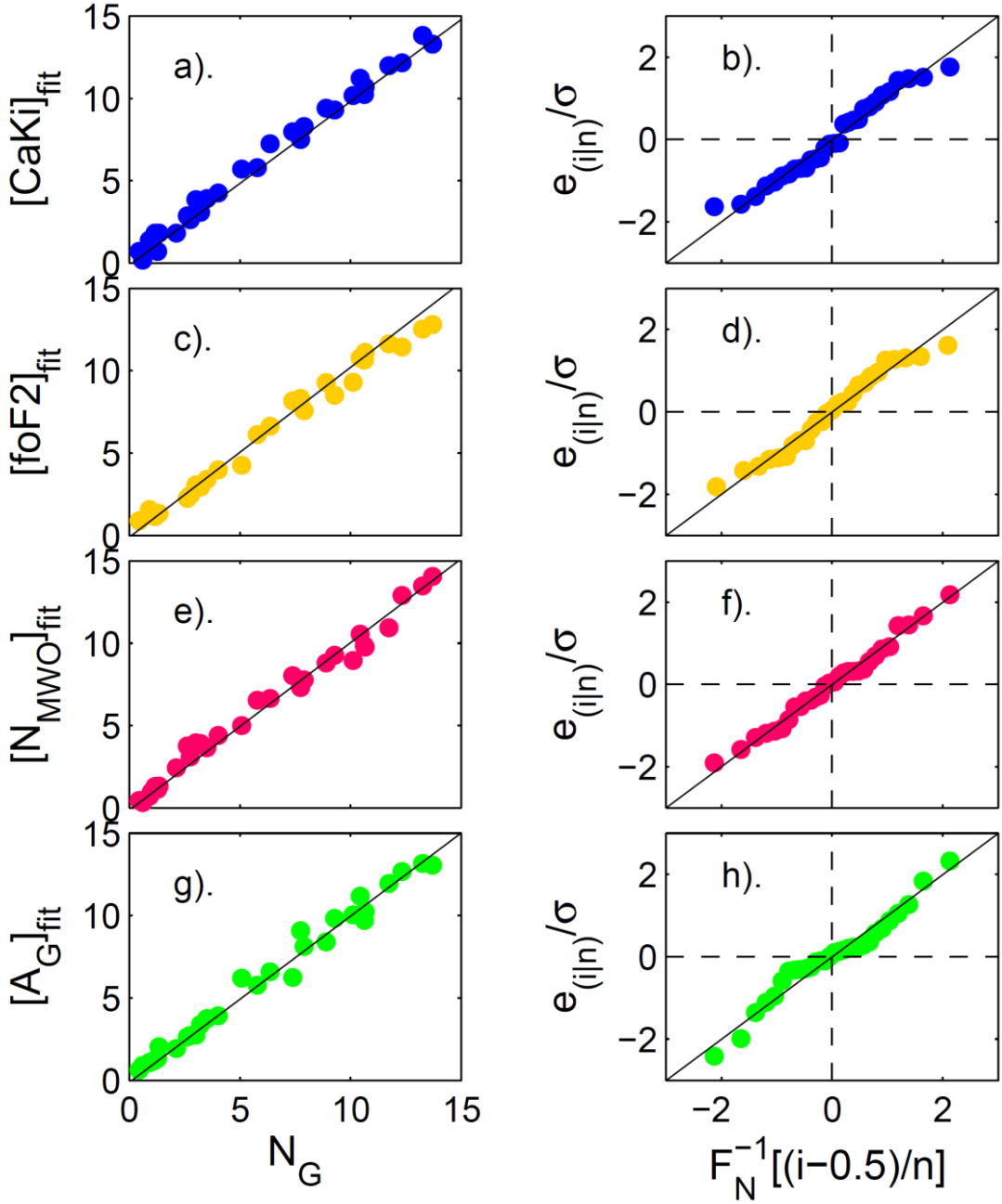

**Figure 4.** Analysis of the regressions between annual means $N_G$ and the other four test series over the interval 1921-1961, as used in figure 2. The left-hand panel (a, c, e and g) show the scatter plots and the best-fit linear regression and the right hand panel (b, d, f and h) the corresponding quantile-quantile (Q-Q) plots in which the ordered standardised fit residuals, $e_{(i|n)}/\sigma$, where $\sigma$ is their standard deviation) are plotted as a function of quantiles of standard normal distribution, $F_N^{-1}[i-0.5/n]$. (a) and (b) are for the Calcium K index, CaKi; (c) an (d) are for the mean dayside Slough F-layer critical frequency, foF2; (e) and (f) are for the MWO sunspot group count, $N_{MWO}$; and (g) and (h) are for the total RGO sunspot area, $A_G$.

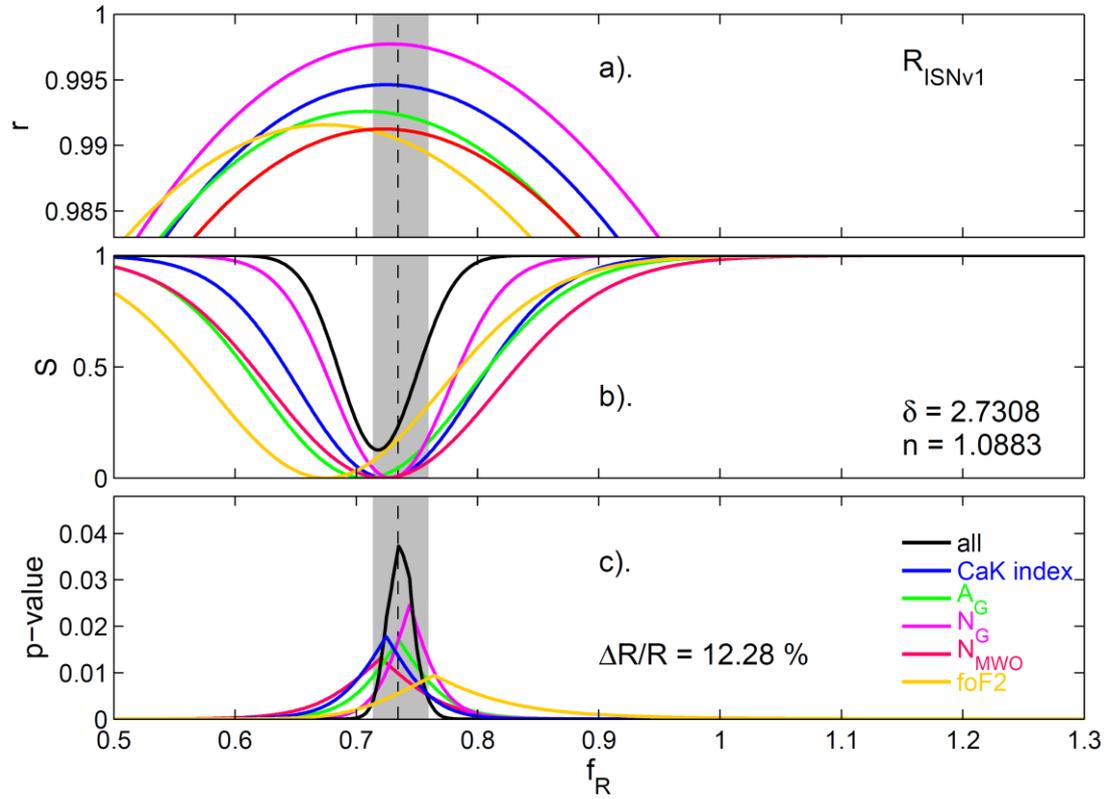

**Figure 5**. Evaluation of the discontinuity around 1946 for version 1 of the Wolf/Zurich/International sunspot number, $R_{ISNv1}$. (a) The correlation $r$ as a function the factor $f_R$ of the adjusted sequence $R_{ISNv1}'$ (generated using equation (1) before 1946 where the tested parameter $R$ is $R_{ISNv1}$) with: (mauve) the RGO group number $N_G$; (green) the corrected RGO total spot area number $A_G$; (blue) the Mount Wilson CaK index; and (orange) the F2 layer critical frequency at Slough, foF2. (b) The significance, $S$ of the differences between the peak $r$ and $r$ at general $f_R$ with (using the same colour scheme). The black line is the combination of the four $S(f_R)$ variations using equation (3). (c) The p-values of the difference in the mean residuals between the "before" (1932-1943) and "after" (1949-1960) intervals, $p(f_R)$, again using the same colour scheme. The black line is the combination of the four pdfs $p(f_R)$ made using equation (2). The vertical dashed line marks the peak, and the grey area the range between the $2\sigma$ points, of the combined $p(f_R)$. The plot is for the optimum offset value $\delta$ of 2.731 and exponent $n$ of 1.088 (see Table 1).

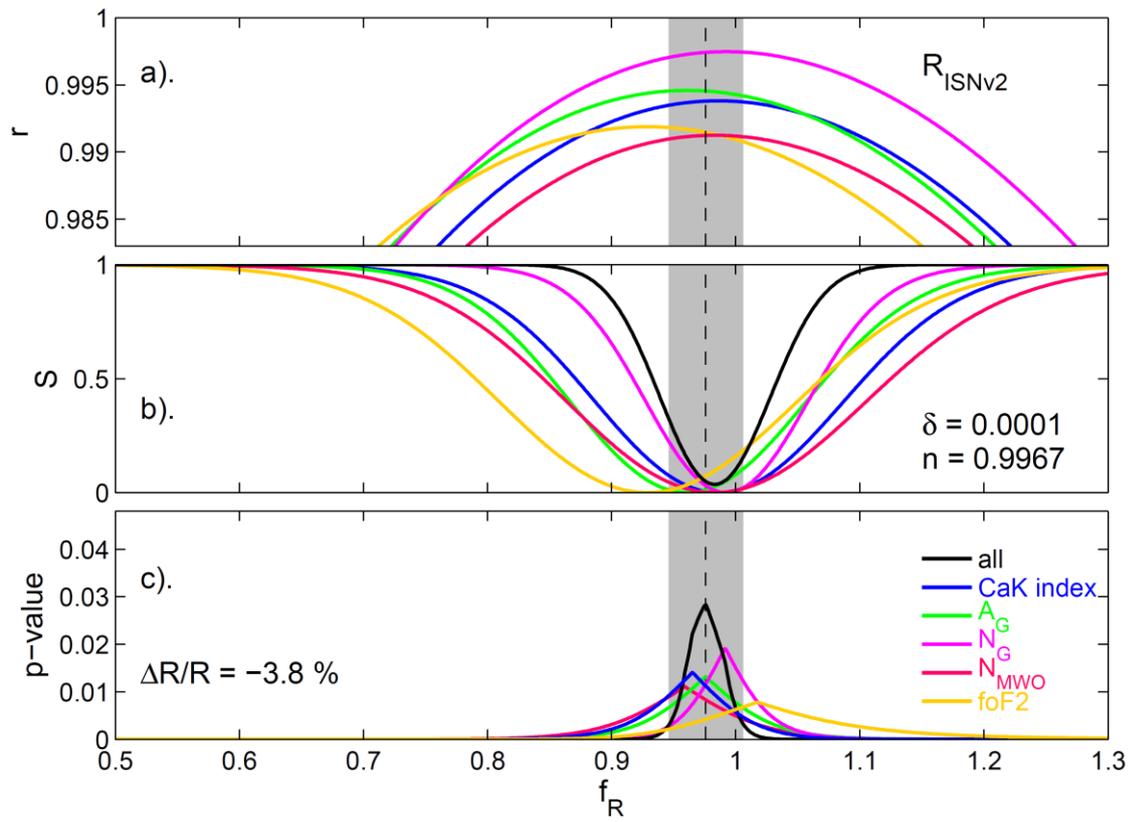

**Figure 6**. Same as figure 5 for version 2 of the of the Wolf/Zurich/International sunspot number, $R_{ISNv2}$.

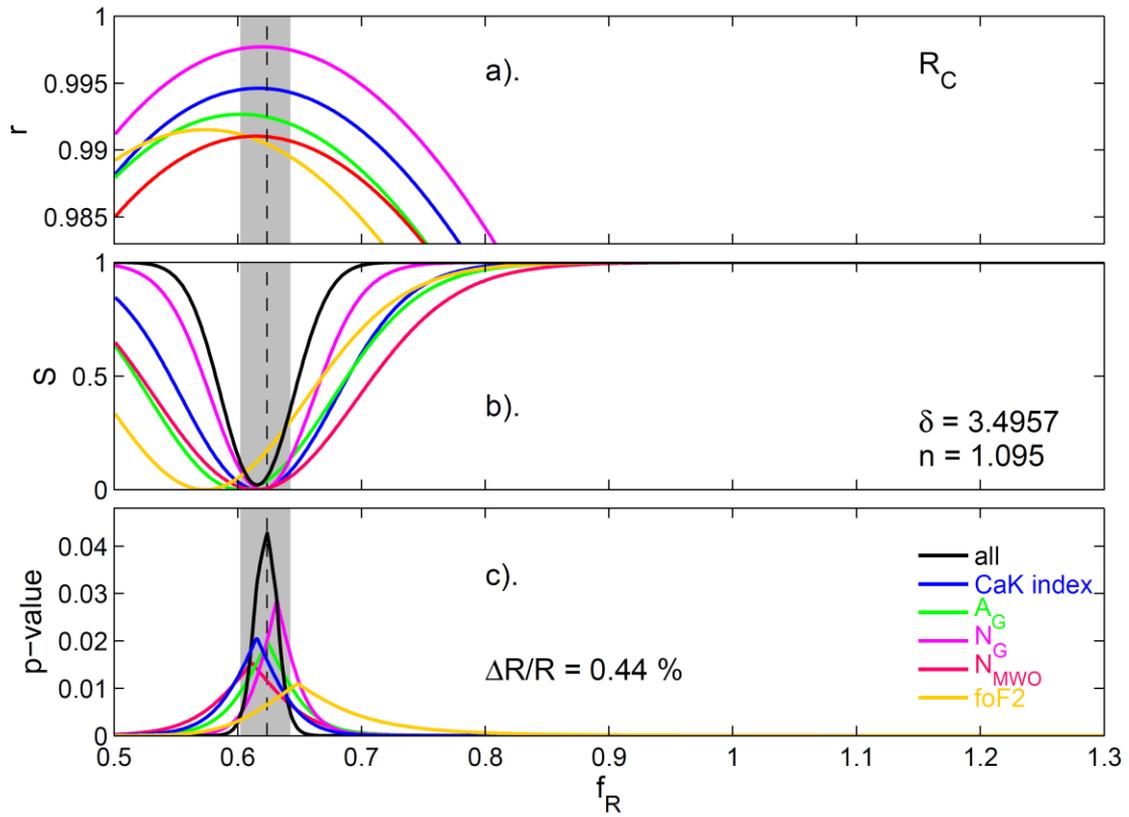

**Figure 7**. Same as figure 5 for the corrected Wolf/Zurich/International sunspot number composite proposed by Lockwood et al. (2015), $R_C$.

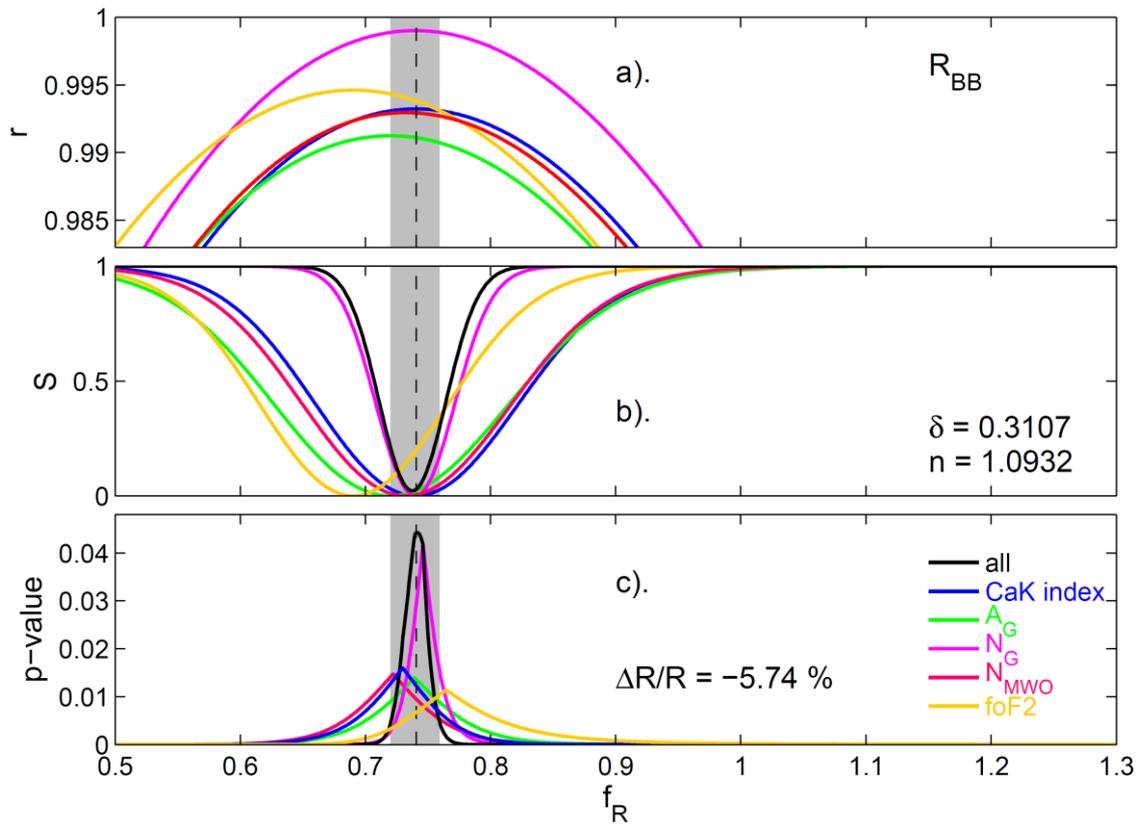

**Figure 8**. Same as figure 5 for the new backbone sunspot group number composite proposed by Svalgaard and Schatten. (2015), $R_{BB}$.

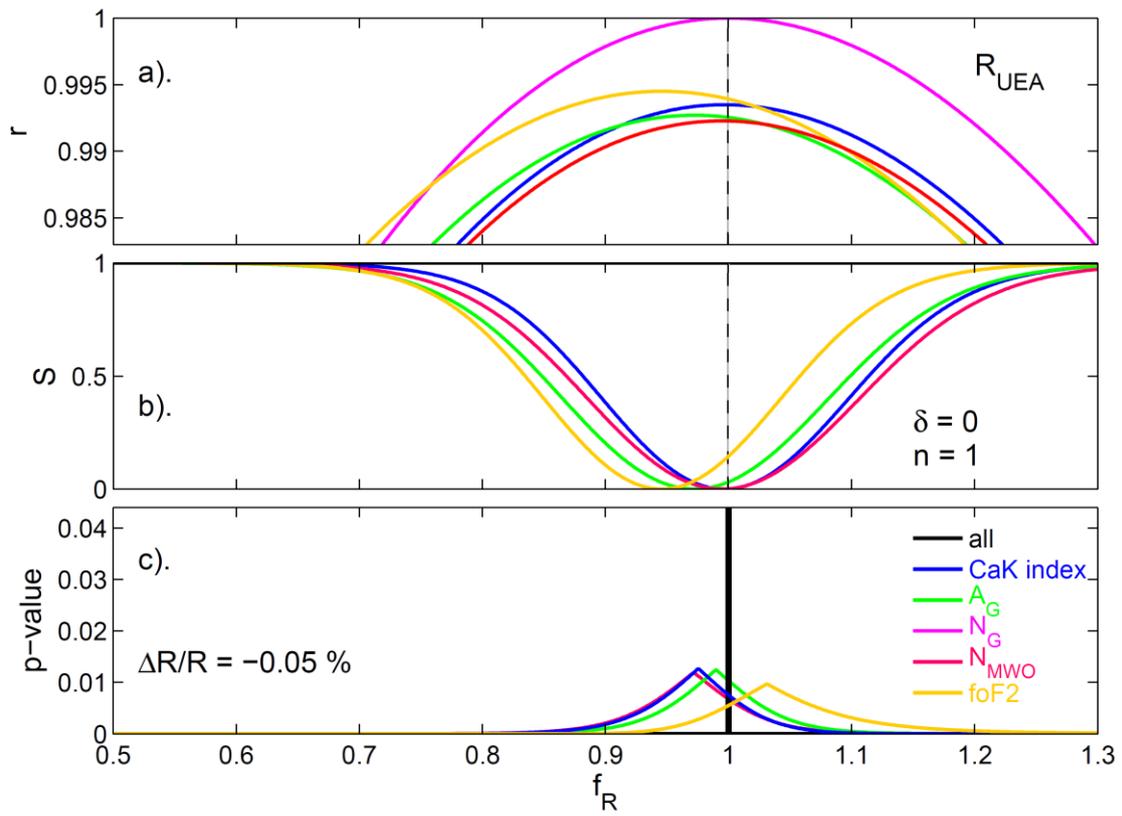

**Figure 9**. Same as figure 5 for the new Usoskin et al. (2015) group number reconstruction. Note that $[p(f_R)]_{N_G}$ and hence $p(f_R)$ are delta functions in this case and although part (c) uses the same $p(f_R)$ scale as figures 3-6, the peak $p(f_R)$ value, $p_m$, is close to unity.

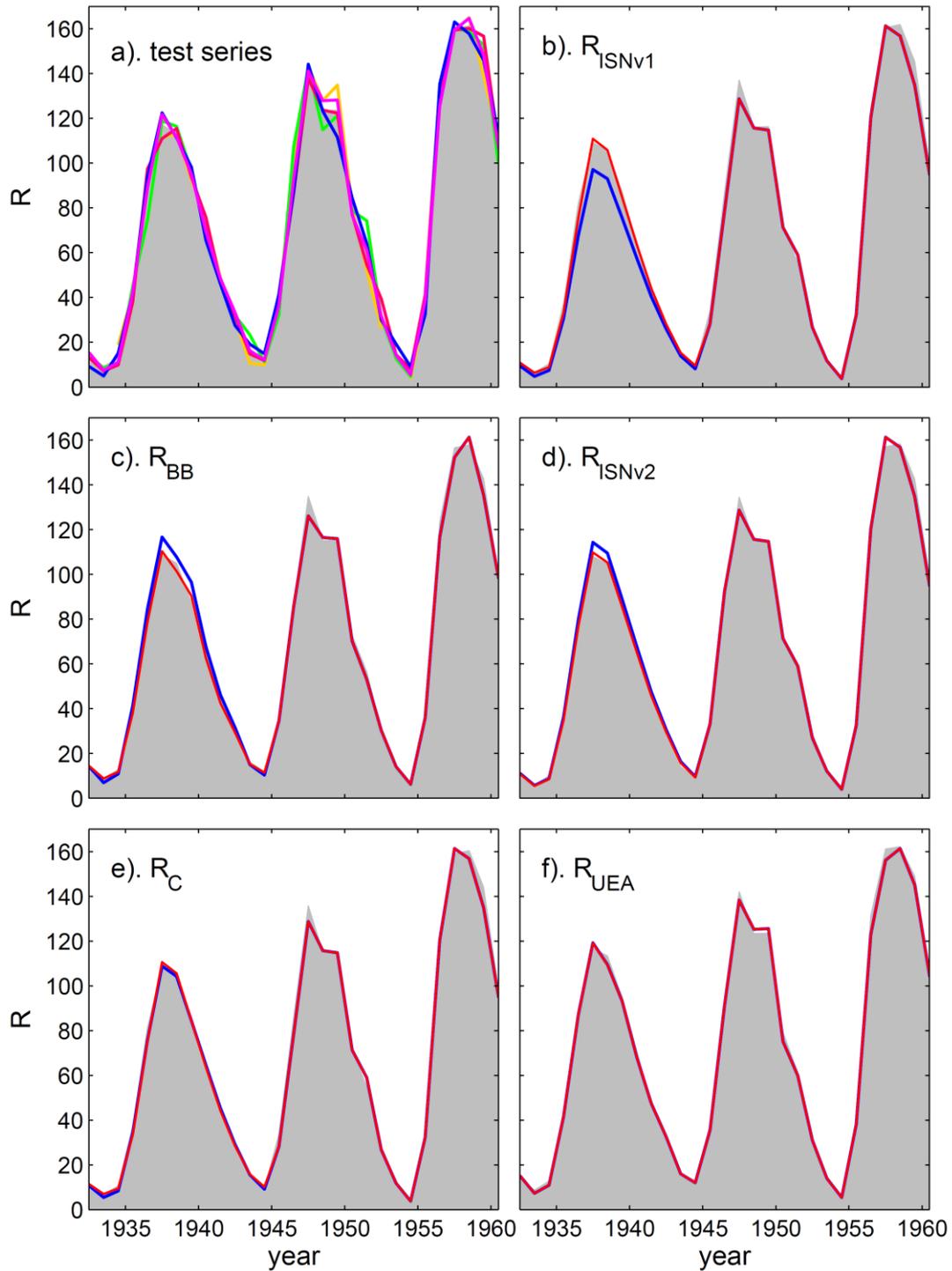

**Figure 10.** Summary of annual mean variations over the optimum test interval 1932−1961. (a) The fitted test series, using the same colour coding as previous figures. The grey shaded area is the mean of the five test series and is repeated in all other parts of the figure. In parts (b)−(f) the blue lines show the original sunspot data series and the red lines the version corrected for before 1946 using the best fits derived in this paper. Because the red lines are plotted second they cover the blue lines where the two agree. The plots are for (b) $R_{ISNv1}$, (c) $R_{BB}$, (d) $R_{ISNv2}$, (e) $R_C$, and (f) $R_{UEA}$. Because this is a mixture of sunspot numbers and sunspot group numbers, all series have been scaled to $R_{ISNv2}$ for the interval 1946−1961.